\documentclass[12pt,a4paper,twoside]{amsart}
 \usepackage{amssymb,latexsym,amsmath,graphicx,epsf,color,hyperref}
\usepackage[table]{xcolor}

 \numberwithin{equation}{section}

\DefineNamedColor{named}{VeryLightOrange}{rgb}{1,0.8824,0.3333}

\def \ba {\begin {eqnarray*}}
\def \ea {\end  {eqnarray*}}

\def \beq {\begin {eqnarray}}
\def \eeq {\end {eqnarray}}

\title{Photographic dataset: random peppercorns}

\author{Teemu Helenius \and Samuli Siltanen}

\begin{document}

\maketitle
\centerline{\today}

\begin{abstract}
This is a photographic dataset collected for testing image processing algorithms. The idea is to have sets of different but statistically similar images. In this work the images show randomly distributed peppercorns. The dataset is made available at \href{www.fips.fi/photographic_dataset.php}{this URL}.
\end{abstract}

\tableofcontents

\clearpage

\section{Introduction}\label{sec:intro}

\noindent This document reports the acquisition, structure and properties of a digital photographic dataset collected at the Industrial Mathematics Laboratory of the Department of Mathematics and Statistics of University of Helsinki, Finland.

The idea is to have objects with the same size scale.

The collected dataset is intended to be ideal for computational approaches based on sparse patch-based dictionaries \cite{Rubinstein2010}. Similar images have already been used in such contexts, see \cite{Soltani2015}.

\section{Materials and Methods}\label{sec:mm}

\subsection{Camera equipment} We use a PhaseOne XF medium-format camera equipped with an achromatic IQ260 digital back. The lens is Phase One Digital AF 120mm F4. 
The pixel size in the resulting 16bit TIFF image file is $8964{\times}6716$.

\subsection{Lighting}

The targets were lit with five Olight X6 Marauder LED flashlights with luminous flux of 5000 lm (nominal value, 4825 lm was measured in the laboratory of the vendor www.valostore.fi). The lights were positioned at roughly equiangular arrangement. The distance of each light from the target was 1,0--1,2m. The lights were heating up quickly as they were used at maximum power. Cooling was enhanced with three regular household fans (Merox Floor Fan FE-45A).

A diffuser was placed between the lights and the target to make the lighting more uniform and to reduce sharp shadows. 

\begin{figure}[ht]
\begin{tabular}{cc}
  \includegraphics[scale=0.36]{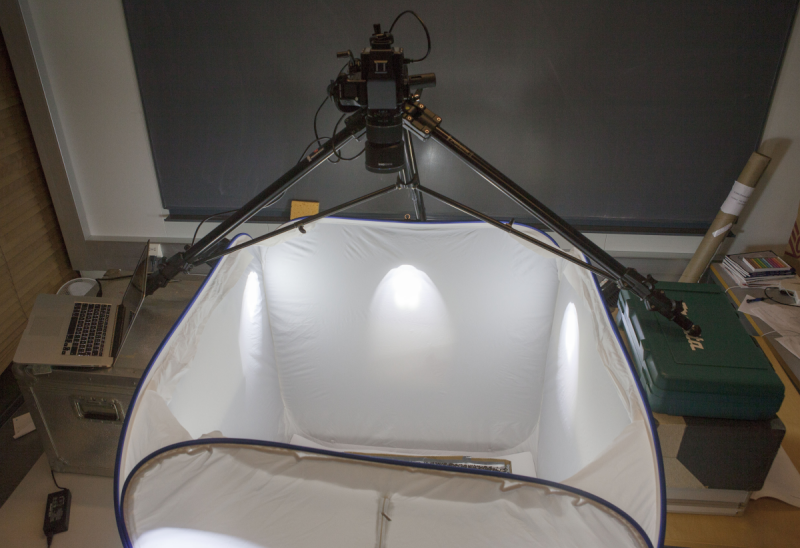} & \includegraphics[scale=0.36]{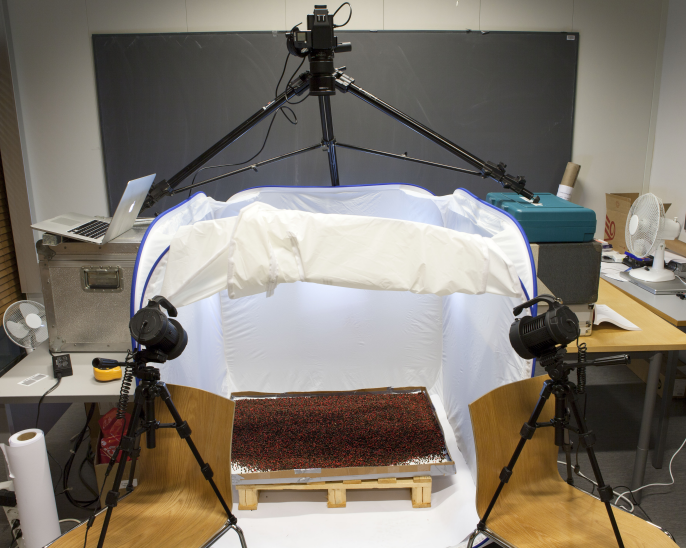}
\end{tabular}

\caption{The imaging setup.}
\label{fig:setup}
\end{figure}

See Figure \ref{fig:setup} for the imaging setup.

\subsection{Size and scale considerations} How to decide how many objects should be visible in the image?

Here are the relevant numbers: the radius of the approximately round peppercorns is 4--5 mm. The patch sizes we have in mind for image processing applications are between $8{\times}8$ (used in JPEG coding) and $20{\times}20$. The full image size is $8964{\times}6716$ pixels.

The relationship between patch size and object size should be chosen wisely. Perhaps it makes sense to have the smallest patch size ($8{\times}8$) roughly equal to the smaller object type, namely 4mm peppercorn. Then using a larger patch would enable having several objects inside one patch. These choices lead to two pixels per millimeter, which means that the full image target area has roughly the size 4,5 meters by 3,4 meters. This leads to an impractically large amount of peppercorns needed to cover the image area.

\begin{figure}[ht]
\includegraphics[scale=0.8]{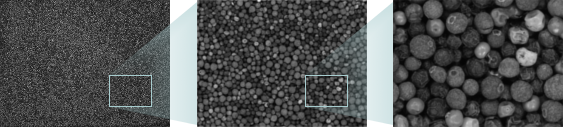}
\caption{The dataset scale.}
\label{fig:zoom}
\end{figure}

However, we cannot predict all possible future uses of this open dataset. It may become important in some study to have even several $8{\times}8$ patches fitting completely inside one peppercorn. Therefore, we choose the scale so that downsampling the images by a factor of four leads to the image of a peppercorn to have diameter of approximately 8 pixels. This means that we need to cover one square meter with objects, see figure \ref{fig:zoom}.

\begin{figure}[ht]
\includegraphics[scale=0.4]{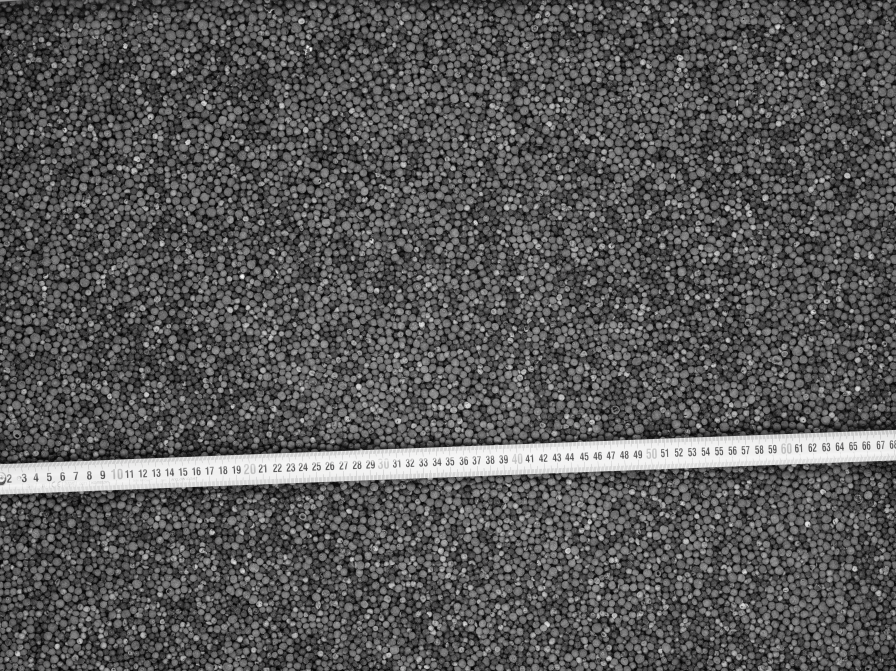}
\caption{The photographed area with a tape measure.}
\label{fig:measured}
\end{figure}

A box of size $1,0 \times 0,7$m was filled with peppercorns, whereas the photographed area was about $0,67 \times 0,5$m, see figure \ref{fig:measured}.

\section{Results}\label{sec:results}

The lens--target distance was 150cm. The image sensor was roughly parallel with the surface of the object layer.
The aperture f-stop was f/11, the ISO setting 200 and the shutter speed was set to $\frac{1}{4} s$.

In total 10 photos were taken as the dataset, see figure \ref{fig:dataset}. Between taking each photo the peppercorns were carefully re-shuffled by hand. Although care was taken to avoid the bottom of the container from being visible through the peppercorns, the possibility of such an occurrence cannot be fully excluded.

\newcommand{\scalefactor}{0.17}

\begin{figure}[ht]
\begin{tabular}{ccc}
  \includegraphics[scale=\scalefactor]{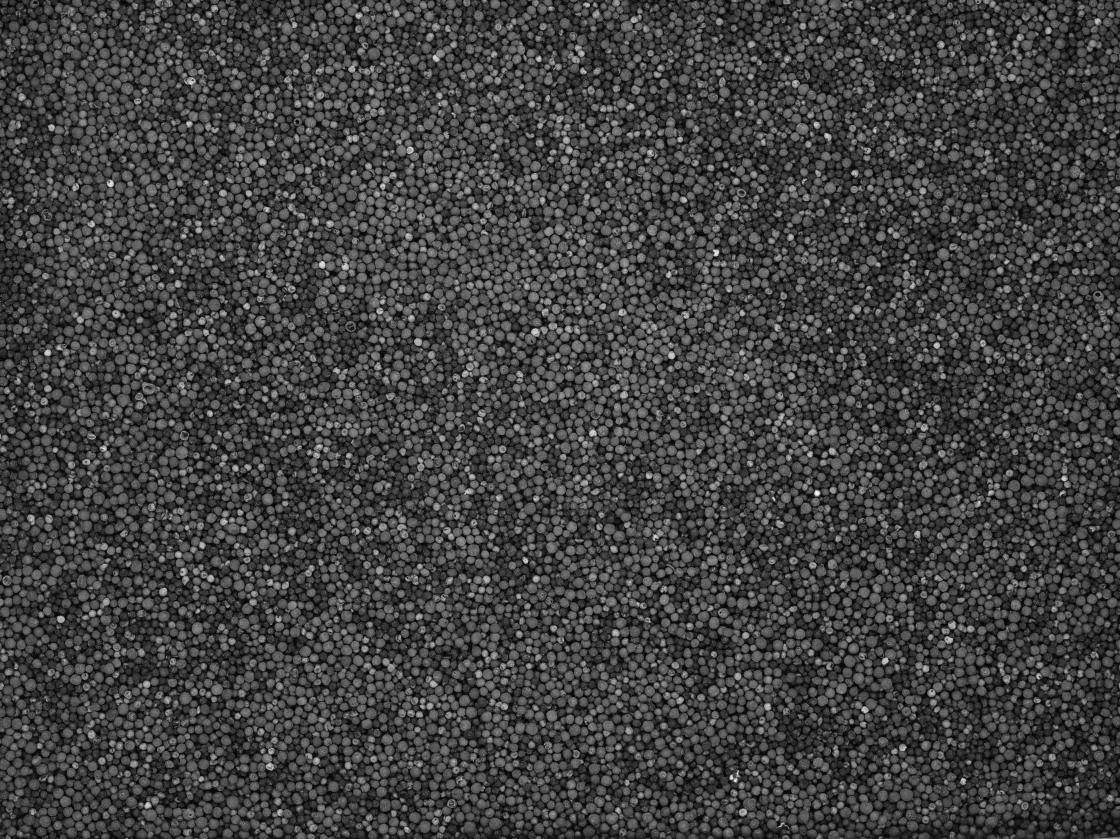} & \hspace{.5cm} & \includegraphics[scale=\scalefactor]{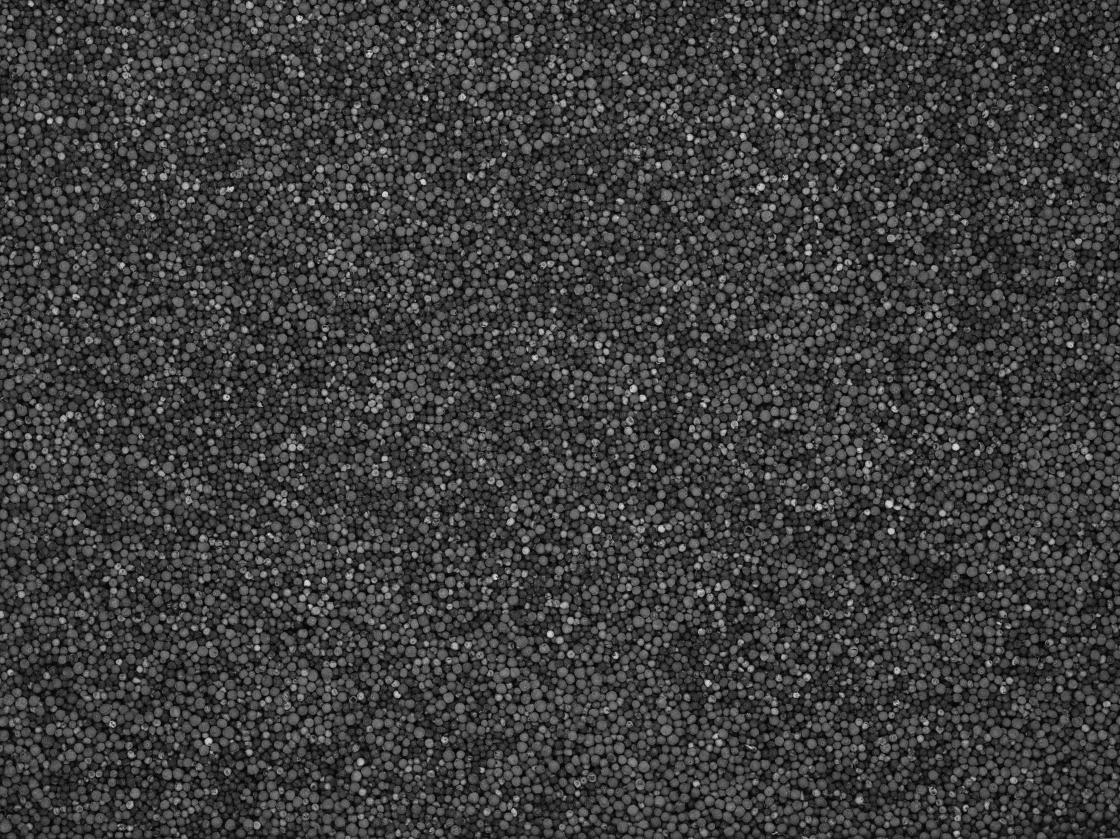} \\
  \includegraphics[scale=\scalefactor]{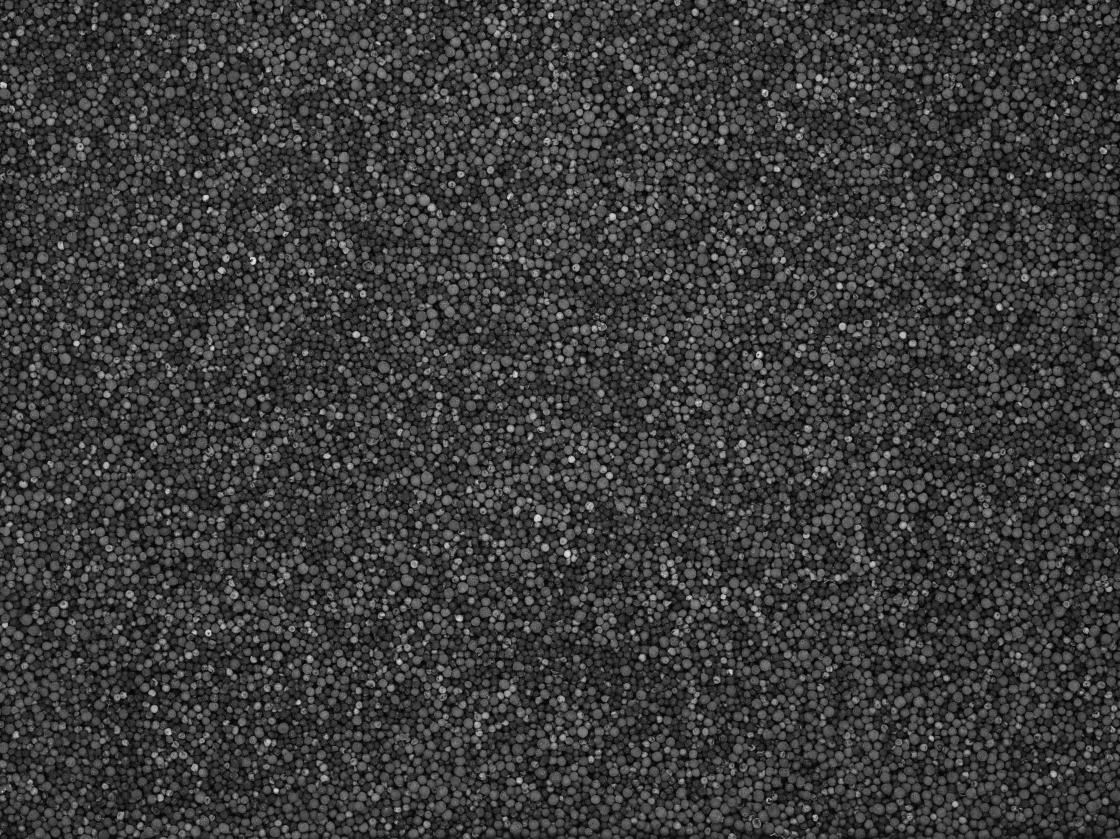} & \hspace{.5cm} & \includegraphics[scale=\scalefactor]{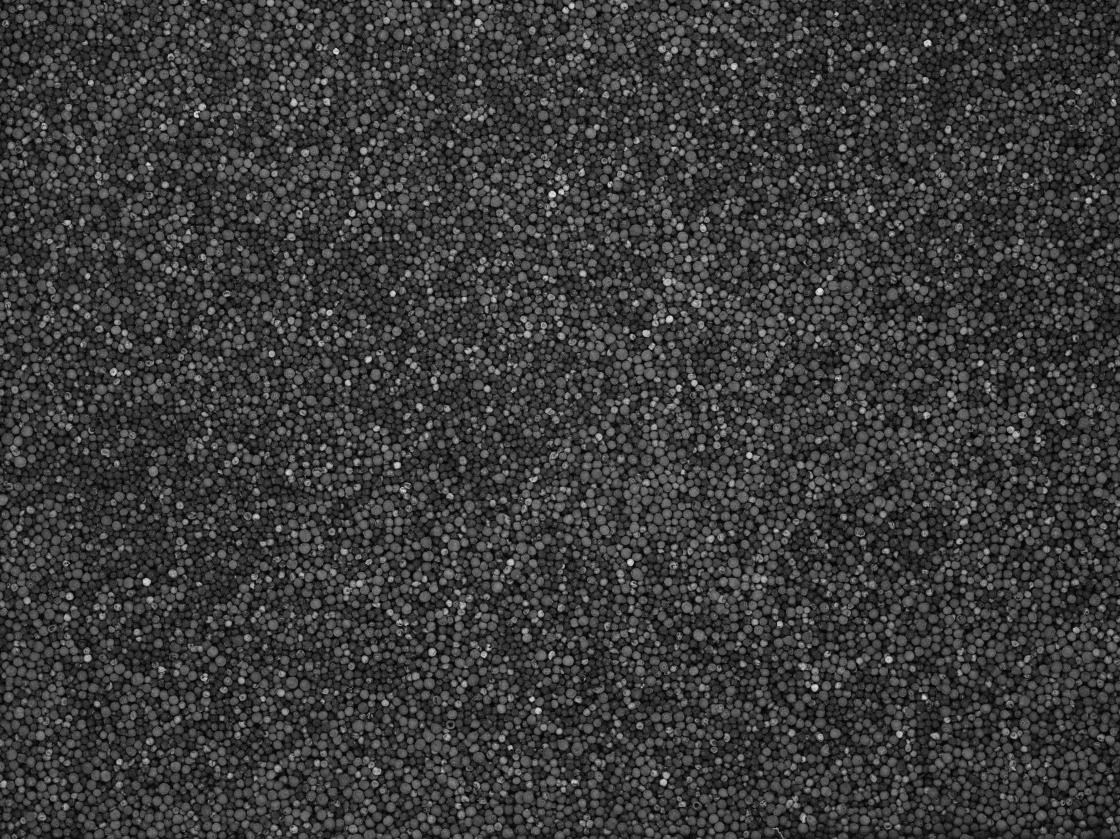} \\
  \includegraphics[scale=\scalefactor]{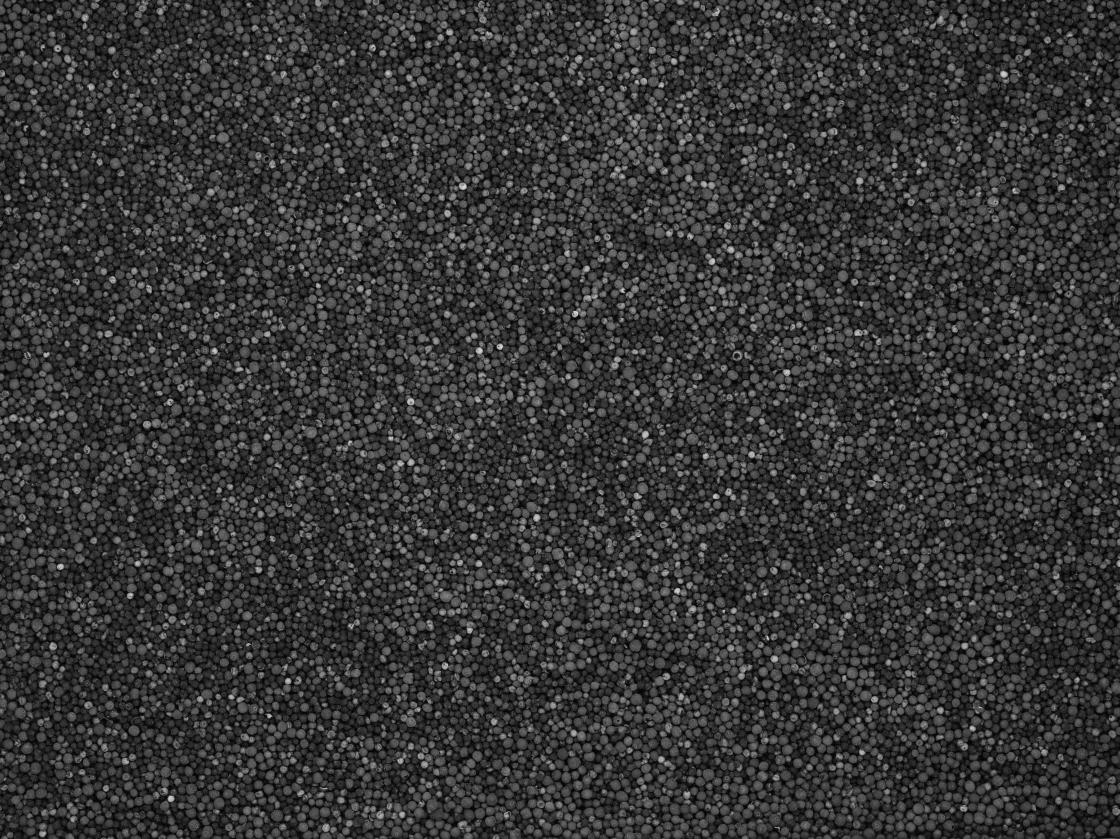} & \hspace{.5cm} & \includegraphics[scale=\scalefactor]{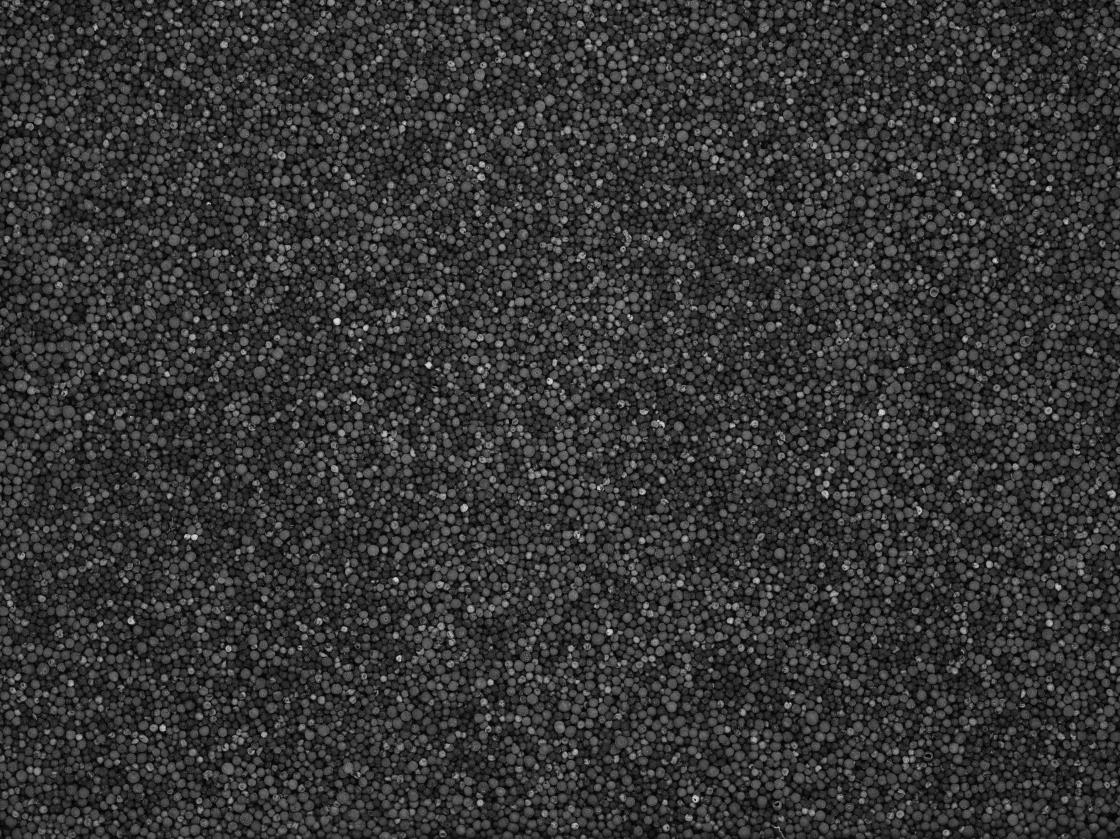} \\
  \includegraphics[scale=\scalefactor]{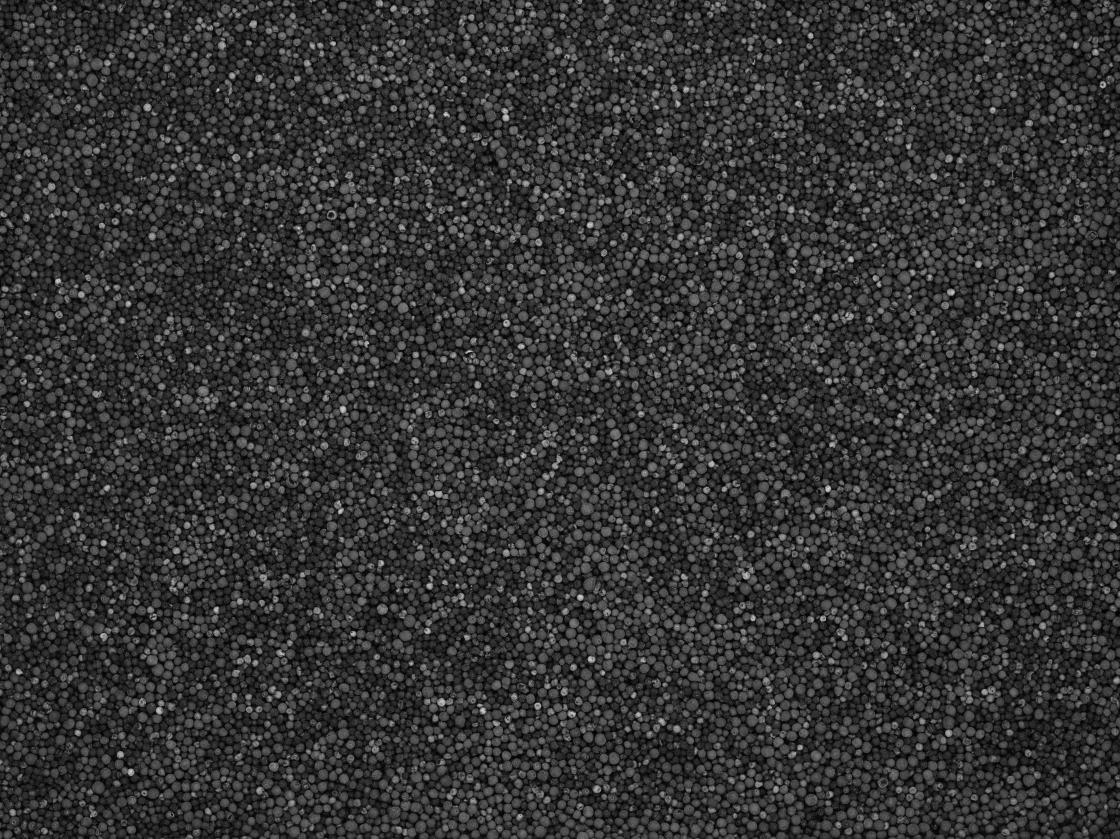} & \hspace{.5cm} & \includegraphics[scale=\scalefactor]{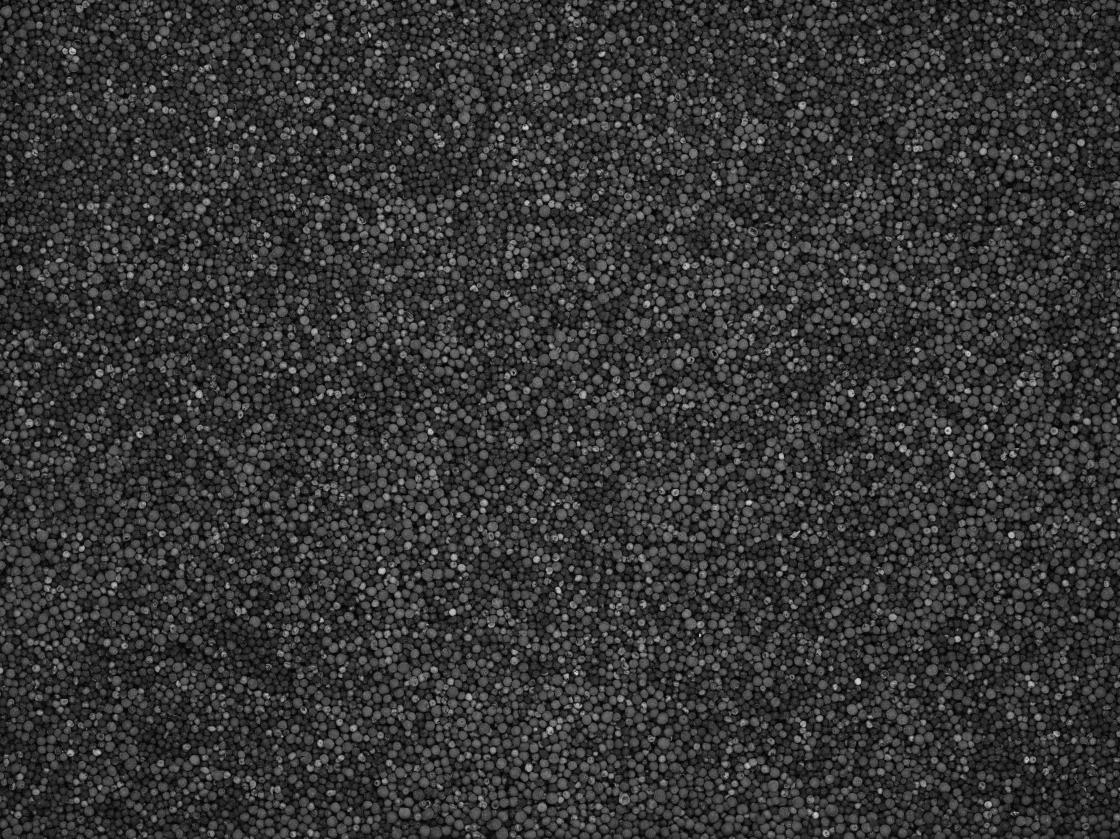} \\
  \includegraphics[scale=\scalefactor]{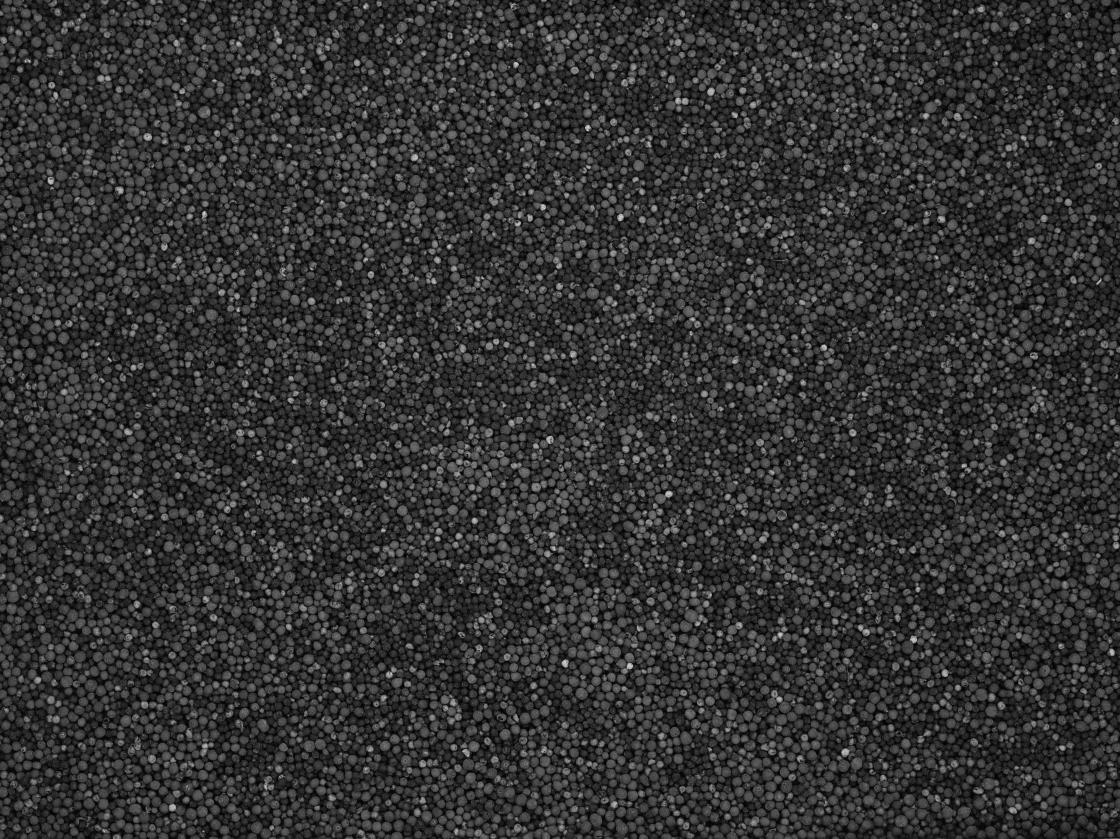} & \hspace{.5cm} & \includegraphics[scale=\scalefactor]{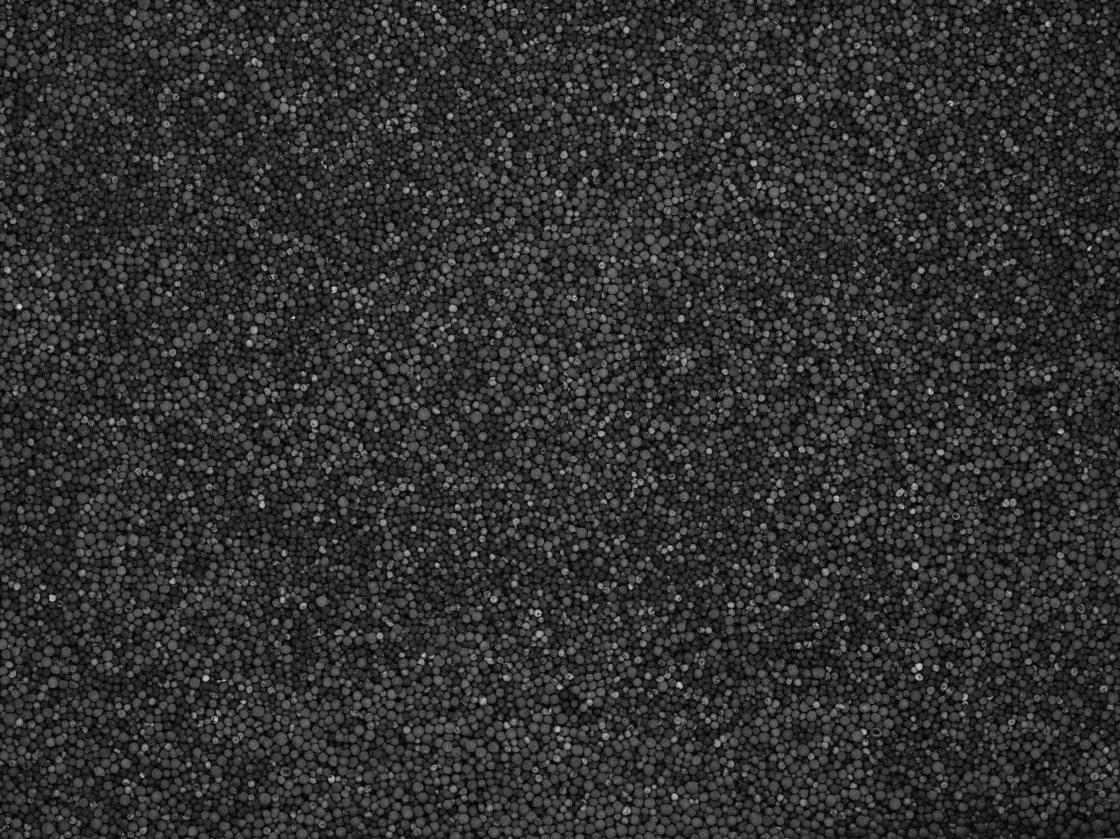}
\end{tabular}

\caption{The dataset downsampled to size $1120 \times 839$.}
\label{fig:dataset}
\end{figure}


\bibliographystyle{siam}
\bibliography{Photodata}

\end{document}